\documentclass{moriond}

\usepackage{color,marvosym, pdfsync}
\usepackage{graphicx, amssymb, epstopdf,url, hyperref,wrapfig,amsfonts,amsmath}
\usepackage[nosort]{cite}
\usepackage[nosort]{cite}
\usepackage[normalem]{ulem}
\usepackage[dvipsnames]{xcolor}

\def\hhref#1{\href{http://arxiv.org/abs/#1}{#1}} 

\newcommand{\gag}{\gamma\gamma}
\newcommand{\TeV}{\,{\rm TeV}}

\newcommand{\GeV}{\,{\rm GeV}}

\def\circa#1{\,\raise.3ex\hbox{$#1$\kern-.75em\lower1ex\hbox{$\sim$}}\,}

\newcommand{\SU}{\,{\rm SU}}
\def\mathscr#1{{\cal #1}}

\newcommand{\beq}{\begin{equation}}
\newcommand{\eeq}{\end{equation}}
\usepackage{url}
\newcommand{\Q}{{\cal Q}}

\newcommand{\pb}{\,{\rm pb}}
\newcommand{\fb}{\,{\rm fb}}

\newcommand{\sfrac}[2]{#1/#2}
\newcommand{\U}{\,{\rm U}}
\newcommand{\X}{\digamma}
\newcommand{\F}{\digamma}
\newcommand{\eq}[1]{~{\rm (\ref{eq:#1})}}
\font\tenrsfs=rsfs10 at 12pt
\font\sevenrsfs=rsfs7 at 10 pt
\font\fiversfs=rsfs5
\newfam\rsfsfam
\textfont\rsfsfam=\tenrsfs
\scriptfont\rsfsfam=\sevenrsfs
\scriptscriptfont\rsfsfam=\fiversfs
\def\mathscr#1{{\fam\rsfsfam\relax#1}}
\newcommand{\Ggg}{\Gamma_{\gamma\gamma}}
\def\Lag{\mathscr{L}}
 \newcommand{\One}{1\hspace{-0.63ex}\hbox{I}}

\newcommand{\Photo}{\vspace{1cm}
\fbox{\includegraphics[height=45mm]{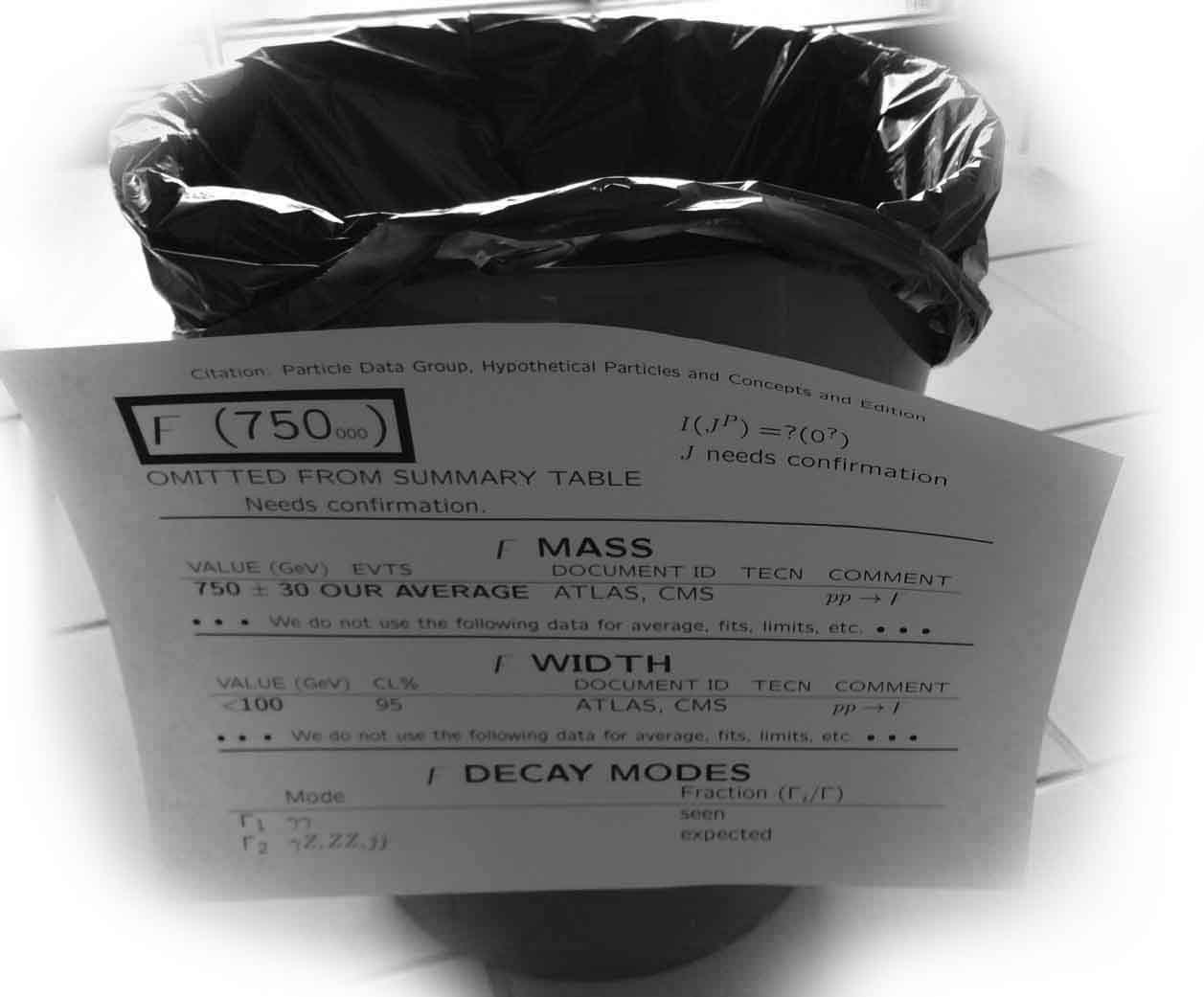}}}

\def\hhref#1{\href{http://arxiv.org/abs/#1}{arXiv:#1}} 
 
\def\art{\@ifnextchar[{\eart}{\oart}}
\def\eart[#1]#2#3#4#5#6{{\rm #2}, {\em #3 \bf #4} {\rm (#6) #5} ({\em #1})}

\makeatletter
\def\article{\@ifnextchar[{\earticle}{\oarticle}}
\def\oarticle#1#2#3#4#5#6{{\rm #1}, {\em ``#6''}, {\rm #2 #3 (#5) #4}}
\def\earticle[#1]#2#3#4#5#6#7{{\rm #2}, {\em ``#7''}, {\rm #3 #4 (#6) #5}  [\hhref{#1}]}
\def\hepart[#1]#2{{\rm #2, \em#1}}
\def\heparticle[#1]#2#3{#2, {\em ``#3''} [\hhref{#1}]}
\def\oarticle#1#2#3#4#5#6{{\rm #1}, {\rm #2 #3 (#5) #4}}
\def\earticle[#1]#2#3#4#5#6#7{{\rm #2}, {\rm #3 #4 (#6) #5}  [\hhref{#1}]}
\def\hepart[#1]#2{{\rm #2}}
\def\heparticle[#1]#2#3{#2, \hhref{#1}}

\begin{document}
\vspace*{4cm}
\title{
Interpreting the 750 GeV digamma excess: a review }

\author{ ALESSANDRO STRUMIA }

\address{CERN, INFN and Dipartimento di Fisica, Universit\`a di Pisa }

\maketitle\abstracts{We summarise the main experimental, phenomenological and
theoretical issues related to  the $750\GeV$ digamma excess.}


The first LHC data about $pp$ collisions at $\sqrt{s}=13\TeV$ agree with the Standard Model (SM), except for a hint of an excess 
in $pp\to \gamma\gamma$  peaked at invariant mass around $750 \GeV$~\cite{seminar}.
We denote the new resonance with the symbol, $\F$,
used in archaic greek as the {\em digamma} letter and later as the number $6 \approx M_\F/M_h$,
but disappeared twice.
New data will tell if the $\F$ resonance disappears or is confirmed.
In the meantime, the $\F$ excess attracted significant theoretical interest~\cite{1512.04850,1512.04913,
1512.04917,1512.04921,1512.04924,1512.04928,1512.04929,1512.04931,big,
1512.04939,1512.05295,1512.05326,1512.05327,1512.05328,1512.05330,1512.05332,1512.05333,1512.05334,1512.05439,1512.05542,1512.05585,1512.05617,1512.05618,1512.05623,1512.05700,1512.05723,1512.05738,1512.05751,1512.05753,1512.05759,1512.05767,1512.05771,1512.05775,1512.05776,1512.05777,1512.05778,1512.05779,1512.05786,1512.05961,1512.06028,1512.06083,1512.06091,1512.06106,1512.06107,1512.06113,1512.06297,1512.06335,1512.06376,1512.06426,1512.06508,1512.06560,1512.06562,1512.06587,1512.06670,1512.06671,1512.06674,1512.06696,1512.06708,1512.06715,1512.06728,1512.06732,1512.06741,1512.06773,1512.06782,1512.06787,1512.06797,1512.06799,1512.06824,1512.06827,1512.06828,1512.06833,1512.06842,1512.06878,1512.06976,1512.07165,1512.07212,1512.07225,1512.07229,1512.07242,1512.07243,1512.07268,1512.07462,1512.07468,1512.07497,1512.07527,1512.07541,1512.07616,1512.07622,1512.07624,1512.07645,1512.07672,1512.07733,1512.07789,1512.07853,1512.07885,1512.07889,1512.07895,1512.07904,1512.07992,1512.08117,1512.08184,1512.08221,1512.08255,1512.08307,1512.08323,1512.08378,1512.08392,1512.08434,1512.08440,1512.08441,1512.08467,1512.08478,1512.08484,1512.08497,1512.08500,1512.08502,1512.08507,1512.08508,1512.08777,1512.08895,1512.08963,1512.08984,1512.08992,1512.09048,1512.09053,1512.09089,1512.09092,1512.09127,1512.09129,1512.09136,1512.09202,1601.00006,1601.00285,1601.00386,1601.00534,1601.00586,1601.00602,1601.00624,1601.00633,1601.00638,1601.00640,1601.00661,1601.00836,1601.00866,1601.00952,1601.01144,1601.01355,1601.01381,1601.01569,1601.01571,1601.01676,1601.01712,1601.01828,1601.02004,1601.02447,1601.02457,1601.02490,1601.02570,1601.02609,1601.02714,1601.03267,1601.03604,1601.03696,1601.03772,1601.04291,1601.04516,1601.04678,1601.04751,1601.04954,1601.05038,1601.05357,1601.05729,1601.06374,1601.06394,1601.06761,1601.07167,1601.07187,1601.07208,1601.07242,1601.07339,1601.07385,1601.07396,1601.07508,1601.07564,1601.07774,
1602.00004,1602.00475,1602.00949,1602.00977,1602.01092,1602.01377,1602.01460,1602.01801,1602.02380,1602.02793,1602.03344,1602.03604,1602.03607,1602.03653,1602.03877,1602.04170,1602.04204,1602.04692,1602.04801,1602.04838,1602.05216,1602.05539,1602.05581,1602.05588,1602.06257,1602.06628,1602.07214,1602.07297,1602.07574,1602.07708,1602.07866,1602.07909,1602.08100,1602.08819,1602.09099,1603.00024,
1603.00287,1603.00718,1603.01204,1603.01377,1603.01606,1603.01756,1603.02203,1603.03333,1603.03421,1603.04248,1603.04464,1603.04479,1603.04488,1603.04495,1603.04697,1603.04993,1603.05146,1603.05251,1603.05592,1603.05601,1603.05668,1603.05682,1603.05774,1603.05978,1603.06566,1603.06962,1603.06980,1603.07190,1603.07263,1603.07303,1603.07672,1603.07719,1603.08294,1603.08525,1603.08802,1603.08913,1603.08932,1603.09350,1603.09354,1603.09550,1604.00728,1604.01008,1604.01127,1604.01640,1604.02029,1604.02037,1604.02157,1604.02371,1604.02382,1604.02803,1604.03598,1604.03940,1604.04076,1604.04822,1604.05319,1604.05328,1604.05774,1604.06180,1604.06185,1604.06446,1604.06948,1604.07145,1604.07203,1604.07365,1604.07776,1604.07828,1604.07835,1604.07838,1604.07941,1604.08307,1605.00013,1605.00037,1605.00206,1605.00542,1605.01040,1605.01898,1605.01937,1605.03585,1605.03831,1605.03909,
1605.04308,1605.04667,1605.04804,1605.04885,1605.04944,1605.05313,
1605.05327,1605.05366,1605.05900,
1605.07183,1605.07962,1605.08183,1605.08411,1605.08681,1605.08736,1605.08741,1605.08772,1605.09018,1605.09359,1605.09647,1606.00557,1606.00865,1606.01052,1606.02716,1606.02956,1606.03026,1606.03067,1606.03097,1606.03415,1606.03804,1606.04131,1606.05163,1606.05326,1606.05865,1606.06375,1606.06677,1606.06733,1606.07082,1606.07171,1606.07903,1606.08753,1606.08785,1606.08811,1606.09592,1607.00204,1607.00810,1607.01016,1607.01464,1607.01910,1607.01936,1607.03829,1607.03878,1607.04187,1607.04276,1607.04534,1607.04562,1607.04643,1607.04857,1607.05403,1607.05592,1607.06074,1607.06234,1607.06440,1607.06608,1607.06900,1607.08517,1608.00153,1608.00382}.
Indeed, unlike many other anomalies that disappeared, the $\gamma\gamma$ excess cannot be caused by a systematic issue, neither
experimental nor theoretical.
Theoretically, the SM background is dominated by tree-level $q\bar q\to\gamma\gamma$ scatterings, which cannot make a $\gamma\gamma$
resonance.\footnote{See~\cite{1605.03909,1606.09592,1607.06440} for attempts of finding a Standard Model interpretation.}
Experimentally, one just needs to identify two photons and measure their energy and direction.
The $\gamma\gamma$ excess is either the biggest statistical fluctuation since decades, or the main discovery.

\section{Data}

During the Moriond 2016 conference
CMS presented new data taken without the magnetic field;
ATLAS presented a new analysis with looser photon selection cuts
(called `spin 2' analysis to distinguish it from the earlier `spin 0' analysis);
furthermore both collaborations recalibrated photon energies in a way optimised around $750\GeV$ rather than around $M_h=125\GeV$.
As a result, the statistical significance of the $\gamma\gamma$ excess increased slightly, both in CMS and in ATLAS.

\begin{figure}[t]
$$\includegraphics[width=0.92\textwidth]{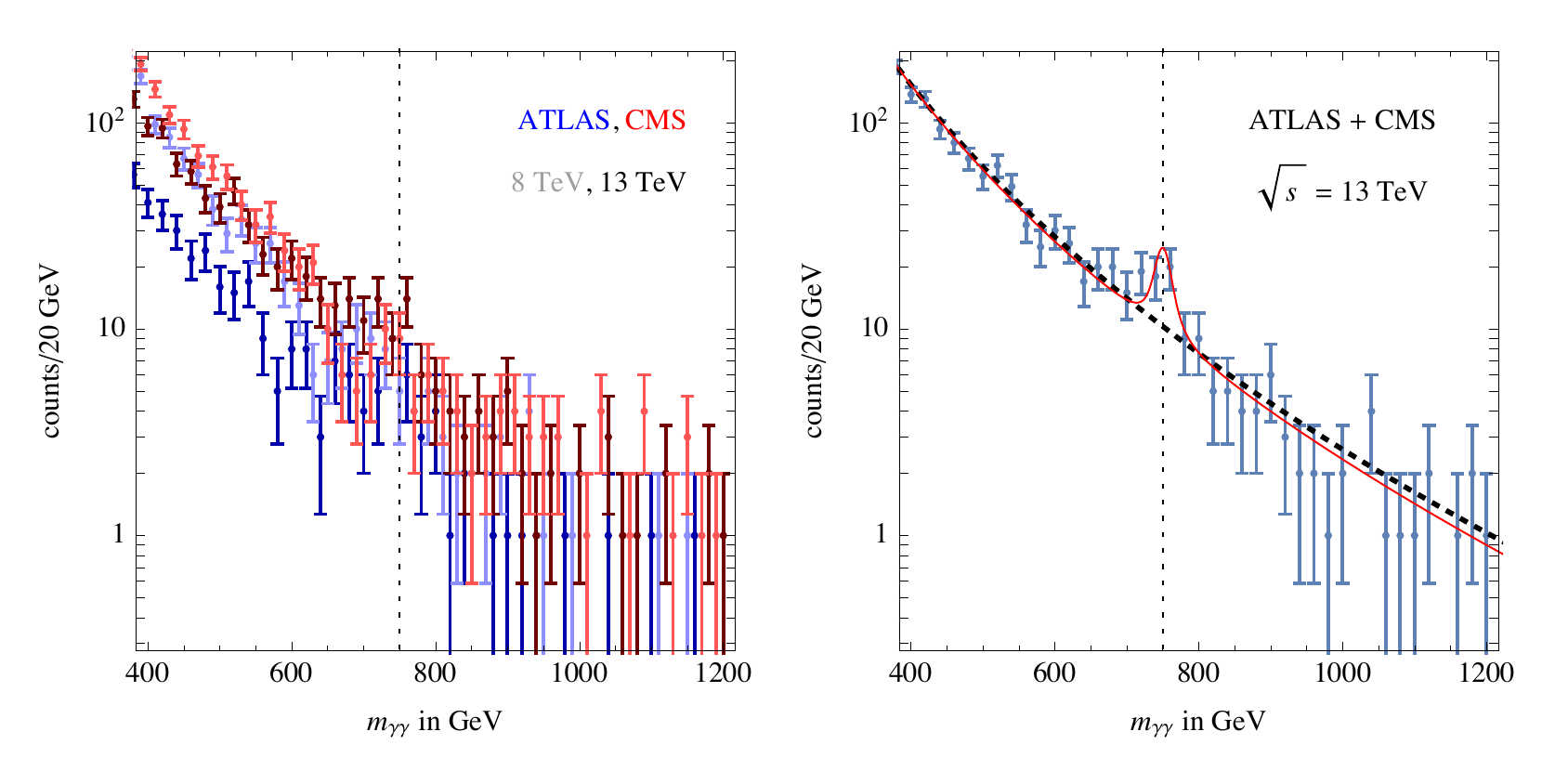}$$
\caption{\em \label{fig:spettri} Left: $\gamma\gamma$ spectra measured by ATLAS (blue) and CMS (red) at 
$\sqrt{s}=8\TeV$ (lighter colors) and $13\TeV$ (darker).
Right: (warning: adult content) spectrum obtained summing ATLAS and CMS counts at $\sqrt{s}=13\TeV$.}
\end{figure}

\begin{figure}[t]
$$\includegraphics[width=0.92\textwidth]{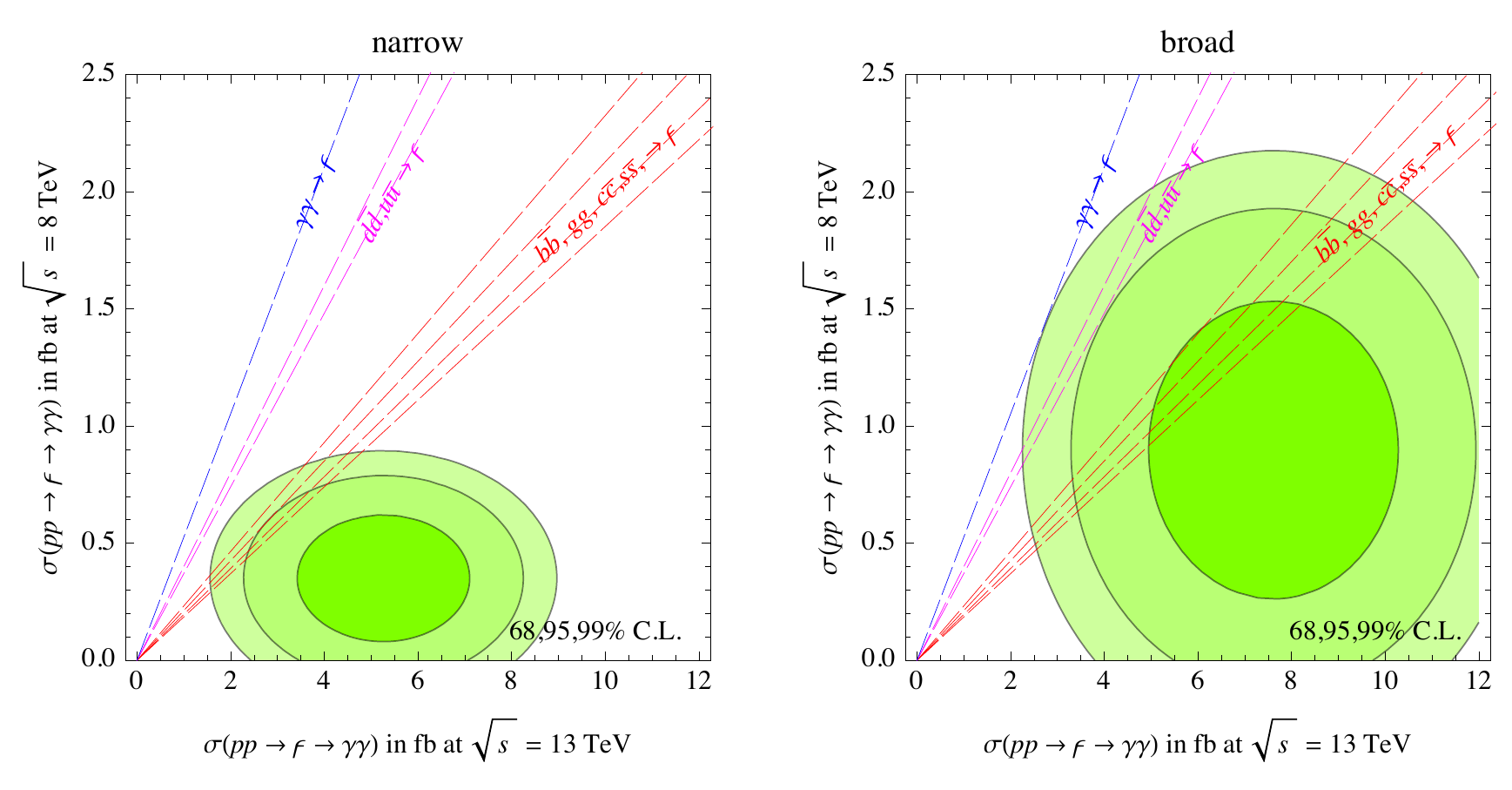}$$
\caption{\em \label{fig:fit} Combination of $pp\to\gamma\gamma$ rates measured by ATLAS and CMS at $750\GeV$.
The diagonal lines show the ratio of $\sqrt{s}=8$ to $13\TeV$ $pp\to\F$ cross sections
predicted for each parton: we see that data favours production from gluons or heavy quarks.}
\end{figure}

Fig.~\ref{fig:spettri}a shows the $\gamma\gamma$ spectra: we consider the `spin 0' ATLAS analysis and the sum of CMS photon categories.
Both ATLAS and CMS find the most statistically significant $\gag$ excess around 750 GeV.
Their consistency can be seen from the peak in  fig.~\ref{fig:spettri}b where
we summed ATLAS and CMS event counts.\footnote{We leave to the intelligence of the reader to evaluate the possible
statistical meaning of this unusual procedure.}
The width of the resonance ranges between $0$ and 100 GeV, and can be larger (`broad') or smaller (`narrow')
than the experimental resolution of about $6-10$ GeV.
The best-fit width is $\Gamma\sim 45\GeV\sim 0.06M_\F$.
The total rates in the two cases, narrow and broad, are:
\beq
\begin{array}{c|cc|cc} 
\sigma(pp\to\F\to\gamma\gamma)  & \multicolumn{2}{c|}{\sqrt{s}=8\TeV} & \multicolumn{2}{c}{\sqrt{s}=13\TeV} \\
& \hbox{narrow} & \hbox{broad}& \hbox{narrow} & \hbox{broad}\\ \hline
\hbox{CMS} & 0.63 \pm 0.31\fb & 0.99\pm 1.05 \fb & 4.8\pm 2.1\fb & 7.7\pm 4.8\fb\\
 \hbox{ATLAS}  & 0.21\pm 0.22\fb & 0.88 \pm 0.46\fb & 5.5\pm1.5 \fb & 7.6\pm 1.9\fb   
 \end{array}\eeq
ATLAS and CMS do not perform a combined analysis.
Na\"{\i}ve combinations of Higgs data gave results  close to the official joint combination, so
fig.~\ref{fig:fit} shows the na\"{\i}ve global fit for $\sigma(pp\to\F\to\gamma\gamma)$ at $\sqrt{s}=8,13\TeV$.
The local excess is about $4\sigma$.
The `Look Elsewhere Effect' (LEE) reduces the
global statistical significance by about $1 \sigma$,
assuming that an excess could have materialised in $\sim 10^2$ other places within the same data-set.
The trial factor can be increased to $10^3$ by considering other similar data-sets,
or to $10^4$ by considering that this search has been repeated about 10 times in the past decades.
We don't need to address such details:
new data will decide if $\F$ will reach the SM scalar $h$ in the Particle Data Group or
if $\F$ will instead reach $N$-rays in the cemetery of anomalies.

\begin{figure}[t]
$$\includegraphics[width=0.96\textwidth]{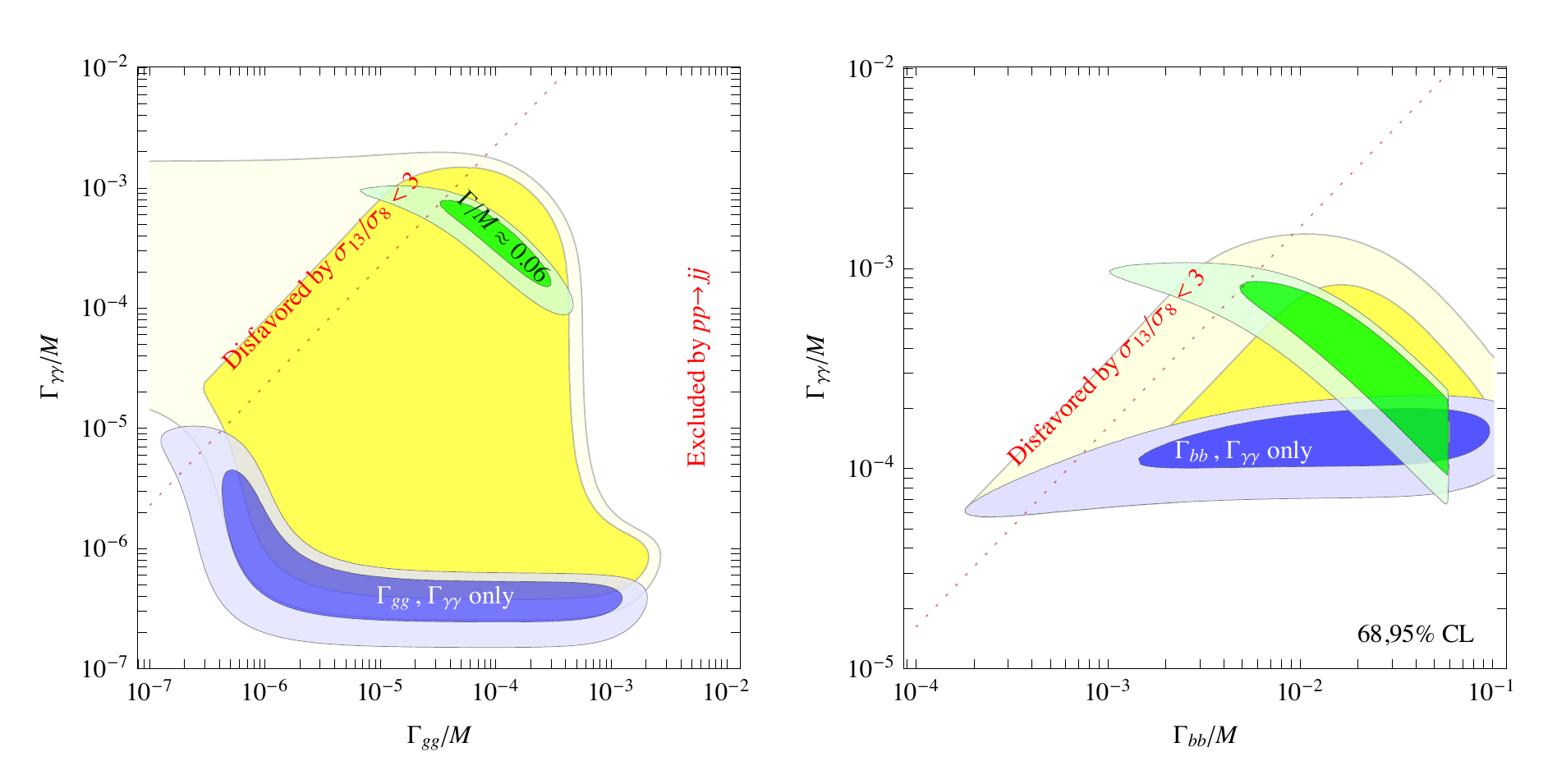}$$
\caption{\em \label{fig:fitGamma} Global fit of the width into $\gamma\gamma, gg$ (left) and $\gag,b\bar b$ (right) 
assuming that diphoton is a scalar with minimal width (blue regions) or $\Gamma/M_\F=0.06$ (green regions)
or a generic width (yellow regions).}
\end{figure}

\section{Widths}
The cross section for single production of a boson $\X$ with spin $J$
can be  written {in the narrow-width approximation} in terms of its decay widths into partons $\wp$,
$\Gamma_\wp=\Gamma(\X\to \wp)$, as
\beq \sigma(pp\to\X ) =\frac{2J+1}{s} 
\sum_\wp C_{\wp} \frac{\Gamma_\wp}{M_{\X}}.
\label{eq:sigmasig}
 \eeq
A resonance with spin $J=1$  is excluded because it cannot decay into $\gamma\gamma$ (Lee-Yang theorem).
The luminosity factors $C_{\wp}$ of the main partons are~\cite{big}
$$\begin{array}{c|cccccccccccc}
\sqrt{s} & C_{b\bar b} & C_{c\bar c} & C_{s\bar s}  & C_{d\bar d}& C_{u\bar u} & C_{gg} 
\\ \hline
8\TeV & 1.07 & 2.7 & 7.2 & 89  & 158& 174  \\
13\TeV  & 15.3 & 36 & 83 & 627 & 1054& 2137
\end{array}.
$$
QCD corrections enhance the cross section by $K_{gg} \simeq1.5$ and $K_{q\bar q}\simeq 1.2$~\cite{big,1603.05978}.
Some authors also consider SM vectors as partons, e.g.\ $C_{\gamma\gamma}\sim 10 ~(60)$ at $\sqrt{s}=8~(13)\TeV$~\cite{1601.00638,1512.05751,big,1601.07187,1604.06446}.
The diagonal lines in fig.~\ref{fig:fit} shows the ratio of cross-section $\sigma_{13}/\sigma_8$ predicted by the various partons:
data disfavor the partons that give the smallest enhancement (light quarks and SM vectors), favouring $\F$ production from heavy quarks or gluons ---
and an even larger $\sigma_{13}/\sigma_8$ enhancement would give a better fit.\footnote{This can be achieved if $pp$ collisions produce some heavier particle
(e.g.\ a heavy vector with mass $\sim 1.5\TeV$) that decays into $\F$ and something invisible~\cite{1512.04928,big,1512.06113,1512.06824,1512.06732,1605.01898,1605.08772},
with a phase space almost closed in order to reproduce the
lack of extra particles and transverse momentum in the $\gag$ excess events.
Such kinematics can be used to fake a large $\F$ width~\cite{1512.08378}.
A large $\F$ width can be faked in other ways:
by having two or more nearby narrow resonances (for example the scalar and pseudo-scalar components of a $\SU(2)_L$ doublet splitted by 
$v^2/M_\F\sim40\GeV$, where $v$ is the Higgs vev)~\cite{big,1601.06374},
or by assuming that $\F$ decays into pairs of light particles with mass $m\circa{<}\GeV$, 
that decay into two or more photons collimated within an angle $\theta\sim m/M_\F$, such that they appear in the detector as
a single $\gamma$~\cite{1512.06083,1512.06671,1512.06833,1602.00949,1603.00024,1605.01898}.
This latter possibility allows to get a large tree-level $\Gamma(\F\to\gag{\rm -like})$ and
can be tested by better studying the $\gamma$ events
(multiple $\gamma$ traveling in the material before the electromagnetic calorimeter
give more $\gamma\to e^+e^-$ conversions than a single photon~\cite{1602.04692}, and
with different kinematical features~\cite{1607.01936});
furthermore it can lead to a displaced vertex, and to no decays into other electroweak vectors.\label{ggjet}}

Fig.~\ref{fig:fitGamma} shows the $\F$ widths that reproduce the $\gamma\gamma$ excess.
In the left (right) panel we 
assumed the production process with the largest (smallest) partonic luminosity, namely $gg$ ($b\bar b$).
The main lesson is that there is a minimal value of $\Ggg$, obtained assuming that
$\F$ has a small width and is dominantly produced from $gg$:
restricting fig.~\ref{fig:fit}a along the $gg$ diagonal line one finds
 \beq \sigma(pp\to\F\to \gamma\gamma)=(2.8\pm0.7)\fb\eeq
such that
\beq
(2J+1) \frac{\Ggg}{M_\F}=\frac{s}{K_{gg}C_{gg}} \sigma(pp\to\F\to\gamma\gamma) = (3.8\pm 0.9)\,10^{-7}\label{eq:-6}\eeq
in agreement with the blue region in fig.~\ref{fig:fitGamma}a.
A larger $\Ggg$ is needed if $\F$ has a larger width (yellow regions) and/or is produced from other partons.
For example, one needs $\Ggg/M_\F\circa{>} 10^{-4}$  (green regions) if $\Gamma/M_\F\sim 0.06$ as favoured by ATLAS.
Finally, reproducing $\sigma(pp\to\F\to\gamma\gamma)$ assuming that production from partonic photons dominates
(a possibility disfavoured by data at $\sqrt{s}=8\TeV$) needs the largest $\Ggg/M_\F\sim 10^{-3}$.

The global fits in fig.~\ref{fig:fitGamma} take into account 
the experimental bounds on other  $\sigma(pp\to\F\to f)$ with final states $f$, as reported in table~\ref{tabounds}.

\begin{table}[t]
$$
\begin{tabular}{c|cc|cc}
final  & \multicolumn{2}{c|}{$\sigma$ at $\sqrt{s}=8\TeV$ }  & \multicolumn{2}{c}{$\sigma$ at $\sqrt{s}=13\TeV$ } \\
state $f$ & observed & expected & observed & expected  \\
\hline  
$e^+e^-  , \mu^+\mu^-$& $<$ 1.2 fb & $<$ 1.2 fb   &  $<5\fb$ & $< 5\fb$ \\
$\tau^+\tau^-$ & $<$ 12 fb & $<$ 15 fb  &  $<60\fb$ & $< 67\fb$  \\
$Z\gamma$ & $<$ 11 fb & $<$ 11 fb  &$<28\fb$ & $< 40\fb$   \\
$ZZ$ & $<$ 12 fb & $<$ 20 fb   & $<200\fb$ & $<220\fb$  \\
$Zh$ & $<$ 19 fb & $<$ 28 fb  &   $< 116\fb$&   $< 116\fb$  \\
$hh$ & $<$ 39 fb & $<$ 42 fb    &   $< 120\fb$&   $< 110\fb$  \\  
$W^+W^-$ & $<$ 40 fb & $<$ 70 fb  &   $< 300\fb$&   $< 300\fb$  \\
$t\bar{t}$ & $<$ 450 fb & $<$ 600 fb  &&\\  
invisible &  $<$ 0.8 pb  &---& 2.2 pb & 1.8 pb \\  
$b\bar b$ & \hbox{$\circa{<} 1\pb$}& \hbox{$\circa{<} 1\pb$}  & & \\  
$ jj$  & $\circa{<}$ 2.5 pb &---  && \tabularnewline   
\end{tabular}\label{eq:bounds}$$
\caption{\em Bounds at $95\%$ confidence level on $\sigma(pp\to \F\to f)$
 cross sections for various final states $f$.
We here assumed  $\Gamma/M_\F\approx 0.06$. 
\label{tabounds}}
\end{table}

\section{Effective Lagrangian}\label{Leff}
So far, $\sigma(pp\to\F)$ and $\F$ decays have been described simply in terms of $\F$ widths.
In order to compute extra related processes we need to make extra theoretical assumptions.
However the Lagrangian interactions of $\F$ differ ---
even in their dimensionality --- depending on the unknown Lorentz and gauge quantum numbers of $\F$: 
\beq \F \stackrel{?}{=}
\begin{pmatrix}
\hbox{spin 0}\cr \hbox{\sout{spin 1}}\cr \hbox{spin 2}\cr\cdots
\end{pmatrix} 
\times
\begin{pmatrix}
\hbox{$\SU(2)_L$ singlet}\cr \hbox{$\SU(2)_L$ doublet}\cr \hbox{$\SU(2)_L$ triplet}\cr\cdots
\end{pmatrix} 
\times
\begin{pmatrix}
\hbox{CP-even}\cr \hbox{CP-odd}\cr \hbox{CP-violating}
\end{pmatrix} \cdots
\eeq
We proceed assuming that $\F$ is a neutral singlet with spin 0, either scalar or pseudo-scalar.\footnote{A spin 2 graviton is disfavoured because 
it couples universally to the conserved energy momentum tensor, such that 
$\sigma(pp\to \F\to e^+e^- + \mu^+\mu^-)=\sigma(pp\to \F\to \gamma\gamma)$,
but no peak is seen in leptons: bounds are reported in table~\ref{tabounds}~\cite{big}.
A spin 2 resonance can be resurrected by assuming that it couples to $\gamma$ more strongly than to leptons;
however this zombie has gauge-dependent cross sections enhanced by inverse powers of $M_\F$
(in effective theories one can restrict to regions of the parameter space where unphysical terms are small).
Angular distributions allow to discriminate spin 0 from spin 2~\cite{1512.06335,1601.07385,1603.08913,1603.09550,1602.02793,1605.09359}.
Bound states with spin 2 which have nothing to do with gravity can couple differently to different particles, and
can have odd parity (while a graviton is even), leading to different angular distributions, which are
as motivated as spin 3 bound states~\cite{1603.04248}.}
Then, the renormalisable interactions of $\F$ are   
\beq
\Lag_4 = \Lag_{\rm SM} + \frac{(\partial_\mu \X)^2}{2}- V(\X,H ) \, ,
\label{lagg4}
\eeq
where
\beq\label{Fpotential}
V(\X ,H) = \frac{m_\X^2}{2} \X^2+\kappa_\X m_\X \X^3 +\lambda_\X \X^4 + \kappa_{\X H} m_\X \X (|H|^2-v^2) +\lambda_{\X H} \X^2 (|H|^2-v^2).
\eeq
This does not give an acceptable $\F\to\gamma\gamma$, so we include 
dimension 5 non-renormalizable interactions, which
are a good approximation to a generic unknown more complete theory with 
extra particles that mediate $\F\to\gamma\gamma$, provided that  such extra particles are much heavier than $M_\F$.
For sure they are much heavier than $M_h$, so we write $\SU(2)_L$-invariant effective operators.
Getting rid of redundant operators, the most generic effective Lagrangian is
\begin{eqnarray}  
\Lag_{5}^{\rm even}&=& \frac{\X}{\Lambda} \bigg[
c_{gg} \frac{{g_3^2}}{2} G^a_{\mu\nu}G^{a\,\mu\nu}+c_{WW}\frac{g_2^2}{2}W^a_{\mu\nu}W^{a\,\mu\nu} +c_{BB} \frac{g_1^2}{2}B_{\mu\nu}B^{\mu\nu}+ c_\psi\left(H
 {\bar \psi}_L \psi_R+{\rm h.c.} \right)
\nonumber
\\
&& 
+  c_H|D_\mu H|^2 -c_H^\prime(|H|^4-v^{4}) \bigg] +\frac{c_{\X3}}{\Lambda}\frac{\X(\partial_{\mu} \X)^2}{2}
\, ,
\label{eq:opsSU2}\label{lagg5}
\end{eqnarray}
for  CP-even $\X$, while
\beq
\Lag_{5}^{\rm odd}= \frac{\X}{\Lambda} \bigg[
\tilde c_{gg} \frac{{g_3^2}}{2} G^a_{\mu\nu}\tilde G^{a\,\mu\nu}+\tilde c_{WW}\frac{g_2^2}{2}W^a_{\mu\nu}\tilde W^{a\,\mu\nu} +\tilde c_{BB} \frac{g_1^2}{2}B_{\mu\nu}\tilde B^{\mu\nu}+\tilde c_\psi\left(i H
 {\bar \psi}_L \psi_R+{\rm h.c.} \right)\bigg].
\label{eq:opsSU2odd}
\eeq
in the CP-odd case.
Operators with higher dimension have been listed in~\cite{1604.06446,1604.07365}.
Couplings to SM fermions $\psi_{L,R}$ are restricted by flavour bounds, which imply that $\F$ can  have large couplings only to pairs of mass eigenstates $t,b,c,\ldots$

\bigskip

\begin{figure}[t]
$$\includegraphics[width=0.92\textwidth]{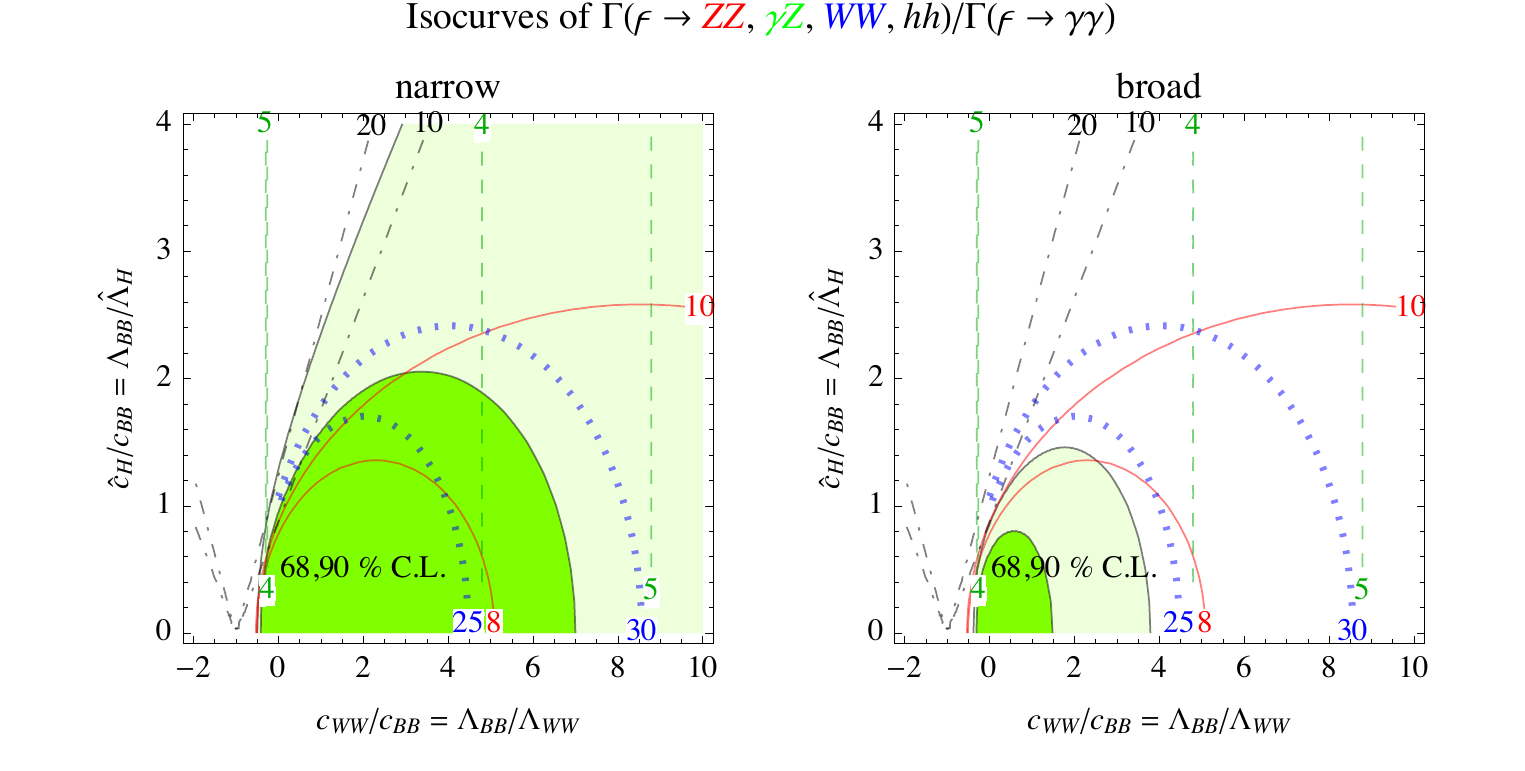}$$
\caption{\em \label{fig:fitc} Best-fit regions at $68,90\%$ C.L.\ (green regions)
for the coefficients of the operators that control $\F$ decays,
assuming that $\F$ is  narrow (left) or broad (right) and produced as $gg\to\F$.
We also show isocurves of $\Gamma(\F\to f)/\Gamma(\F\to\gamma\gamma)$
for $f=ZZ$ (red), $f=\gamma Z$ (green dashed), $WW$ (blue dotted), $hh$ (black dot-dashed).}
\end{figure}

Defining  ${c_{\gamma\gamma}} = {c_{BB}}+ {c_{WW}}$, the above  effective Lagrangian leads to the decay widths
\beq
\Gamma_{ \gamma\gamma}= \frac{\pi \alpha^2 M^3_\X}{\Lambda^2}(c_{\gamma\gamma}^2+\tilde c_{\gamma\gamma}^2)\,,\qquad
\Gamma_{gg}= K_{\Gamma_{gg}} \frac{8 \pi \alpha_3^2 M^3_\X}{\Lambda^2} (c_{gg}^2+\tilde c_{gg}^2)\,,\eeq
($K_{\Gamma_{gg}} =1.35$ when the other couplings are renormalised at $\bar\mu=M_\F$)
together with other decays modes with characteristic rates
$$
\begin{array}{ll}\displaystyle
\Gamma_{ h h}  = \frac{M_{\X}^3}{128 \pi  \Lambda^{2}}\hat c_{H}^2 , &
\displaystyle
\Gamma_{ZZ}= \frac{\pi\alpha^{2}M_{\X}^{3}}{\Lambda^{2} s_{\rm W}^{4}c_{\rm W}^{4}}\left(c_{ZZ}^{2} +\tilde{c}_{ZZ}^{2}\right)+\Gamma_{ h h},\\
\displaystyle
\Gamma_{\gamma Z}= \frac{2 \pi \alpha^{2} M_{\X}^{3}}{s_{\rm W}^{2} c_{\rm W}^{2} \Lambda^{2}}\left(c_{\gamma Z}^{2}+\tilde{c}_{\gamma Z}^{2}\right),&
\displaystyle
 \Gamma_{WW}=\frac{2\pi\alpha^{2}M_{\X}^{3}}{\Lambda^{2} s_{\rm W}^{4}}(c_{WW}^{2} +\tilde{c}_{WW}^{2})+2\Gamma_{ h h},
 \end{array}$$
where
${c_{\gamma Z}} = s_{\rm W}^2{c_{BB}}- c_{\rm W}^2{c_{WW}}$,
${c_{ZZ}} = s_{\rm W}^4{c_{BB}}+c_{\rm W}^4{c_{WW}}$ and
$\hat c_H  =  c_{H} +2 \kappa_{\X H} \Lambda/M_\X$.
We neglected terms suppressed by $M_{W,Z,h}/M_\F$, computed in~\cite{1604.06446}.

Experiments only set upper bounds on these extra decay modes.
The bounds are satisfied assuming, for example, that $\F$ coupes  to hypercharge only.
A $\F$ coupled  to $\SU(2)_L$ vectors only is instead  disfavoured by bounds on $\F\to WW,ZZ$, especially if $\F$ is broad.
The limiting values are
$$\begin{array}{c|ccc}
\hbox{operator} &
\Gamma_{Z\gamma}/\Ggg &
\Gamma_{ZZ}/\Ggg&
\Gamma_{WW}/\Ggg \\ \hline
c_{WW}~\hbox{only} & 
\displaystyle{2}/{\tan^2\theta_{\rm W}}\approx 7 &
\displaystyle {1}/{\tan^4\theta_{\rm W}}\approx 12 &
 \displaystyle {2}/{\sin^4\theta_{\rm W}}\approx 40^{\phantom{1^1}}\\[1mm] 
c_{BB}~\hbox{only} &2 \tan^2\theta_{\rm W} \approx 0.6   & \tan^4\theta_{\rm W}\approx 0.08 & 0\\
\end{array}\ .$$
Fig.~\ref{fig:fitc} shows the best-fit region in the $(c_{WW}/c_{BB}, \hat c_H/c_{BB})$ plane.
Future observations of extra decay modes will over-constrain the fit.
In particular, the $\F$ width into $\gamma Z$ {\em or} the width into $ZZ$ can be fine-tuned to zero, but not both:\footnote{A resonance
that apparently decays only to $\gag$ is possible if one or both photons 
actually are collimated jets of photons: models that realise this have been presented in footnote~\ref{ggjet}.}
the diphoton cannot be only a diphoton.
The minimal extra effect $\Gamma_{\rm extra}>0.28 \Gamma_{\gamma\gamma}$ is obtained for $c_{WW}\approx 0.04 c_{BB}$.


\medskip

Concerning the decays into Higgs components (the physical $h$, and the 3 Goldstones eaten by the $W^\pm$ and the $Z$),
they can be equivalently described in terms of a mixing angle $\theta_{h\F} $
between the Higgs mass eigenstate $h$ and the diphoton
\beq \label{mixingangle} \tan2\theta_{h\F} = \frac{2v(m_{\X} \kappa_{\X H}+\sfrac{c_{H}^{\prime}v^2}{\Lambda})}{m_\X^2 - m_H^2}.\eeq
The experimental bound $\Gamma(\X\to ZZ)\circa{<} 20\, \Gamma(\X\to\gamma\gamma)$ implies that such angle is small
\beq |\sin\theta_{h\F} | \circa{<} 0.015 \sqrt{\sfrac{\Gamma(\X\to\gamma\gamma)}{10^{-6}M_\X}}  .\eeq

\section{The everybody's model}
Many renormalizable models can realise one or more of the effective operators: for example couplings to fermions can be mediated at tree level by
one extra Higgs doublet or by extra vector-like fermions.
A  different class of models attracted most attention: those where
the new particles that mediate $\F\to\gamma\gamma$ (the only process mandated by data) 
also mediate $gg\to\F$.
Such renormalizable models are obtained adding to the SM a neutral scalar $\F$ and 
extra charged fermions or scalars $\Q$ in order to mediate
the effective $\F$ couplings to vectors in the following way:
\beq\raisebox{-1.5cm}{ \includegraphics[width=0.34\textwidth]{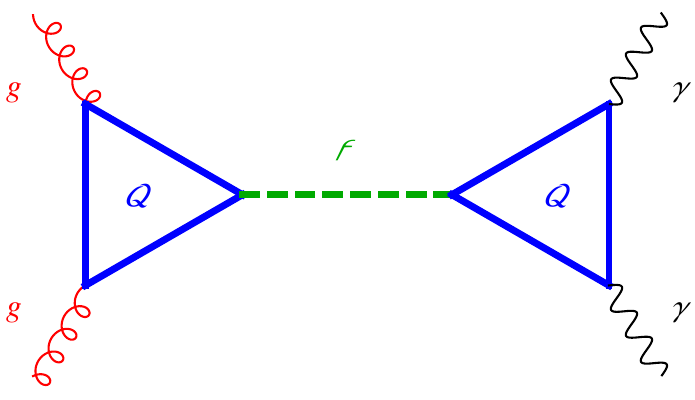}}\label{eq:Feyn1}\eeq
The role of $\Q$ cannot be played by SM particles, because $\F$ would also decay into them at tree level, 
violating the bounds in eq.\eq{bounds}.
For example, a coupling of $\F$ to top quarks would contribute as $\Ggg\sim 10^{-5}\Gamma_{t\bar t}$.

The resulting decay widths are
\beq
\frac{\Gamma_{gg}}{M_\F} = \frac{\alpha_3^2}{2\pi^3}X_{gg}^2\sim
7.2\times 10^{-5} X_{gg}^2,\qquad
\frac{\Ggg}{M_\F} = \frac{\alpha^2_{\rm em}}{16\pi^3}X_{\gag}^2\sim5.4~10^{-8}
X_{\gag}^2
\label{eq:VM}
\eeq
where $X_{gg}\sim N y M_\F/M_\Q$ and
$X_{\gamma\gamma} \sim N q^2 y M_\F/M_\Q$ are loop functions that
only contain model-dependent order one factors: the multiplicity $N$ of states $\Q$, their
color or electric charge $q$, the strength $y$ of their coupling to $\F$, and their mass.
We do not report here the well known expressions for the loop factors, see e.g.~\cite{big}.
If $\F$ is a pseudo-scalar, the fermion loop is resonantly enhanced 
when $M_\Q=M_\F/2$~\cite{1603.04464,1603.07263,1512.06715}.
A useful general result holds if $\F$ is a scalar much lighter than $\Q$: 
the $\F$ coupling to vectors are determined
by the contribution $\Delta b^\Q$ of particles $\Q$ to the gauge beta functions as
\beq\label{eq:LET}
 \Lag_{\rm eff} = \sum_{i,\Q} 
   \Delta b_{i}^\wp \frac{\alpha_i}{8\pi} (F_{\mu\nu}^{i})^2 \ln\frac{M_\Q(\F)}{M_\Q}
   \eeq
where  $M_\Q(\F) $ is the $\Q$ mass for a generic vev of $\F$.
It can be computed in any given model, and expanding it at first order in $\F$,
$ \ln\sfrac{M_\Q(\F)}{M_\Q}\simeq\F/v_\Q$,
gives the desired $\F$ coupling to vectors in terms of the model-dependent constant $v_\Q$.
The main message is that
\begin{quote}
order one charges, multiplicities and couplings in eq.\eq{VM} 
can reproduce the value of $\Ggg/M_\F\sim 10^{-6}$ suggested by the measured
$\sigma(pp\to\F\to\gamma\gamma)$ {\em assuming that the diphoton is narrow}.
\end{quote}
The conclusion drastically changes if instead the diphoton is broad.
ATLAS gives a $\sim1\sigma$ hint in favour of $\Gamma/M_\F\sim 0.06$.
Such a large width, by itself, would not be a problem: it can be obtained
as a tree level two-body decay with a order one coupling,
analogous to the top Yukawa coupling.
The bounds on $\F$ decays of table~\ref{tabounds} allow for a large $\F$ decay width into jets and/or invisible channels, such as neutrinos or Dark Matter.
However, if $\F$ is broad, larger values of $\Ggg/M_\F \sim 10^{-3-4}$ are needed to reproduce $\sigma(pp\to\F\to\gamma\gamma)$.
This is the problem: according to eq.\eq{VM}, in order to reproduce such a large $\Ggg$, $y$ and/or $q$ and/or $N$ need to be large.
Apart from plausibility issues, all these possibilities lead to some coupling becoming non-perturbative and hitting a
Landau pole at energies not much above the diphoton mass~\cite{1512.07624,1512.08307,1512.08500,big,1601.02447,1602.01460,1602.03653,1602.04170}.
A large electric charge $q\sim 3$ or $N\gg 1$ states with $q\sim 1$ imply a fast running of the hypercharge gauge coupling\footnote{This can be
experimentally tested trough high-energy tails of $pp\to\ell^+\ell^-$ distributions at LHC~\cite{1602.03877,1602.04801},
which probe the SM electroweak couplings renormalised at $m_{\ell\ell}\sim 2\TeV$,
and trough precision measurements around the $Z$-peak at a future $e^+e^-$ collider.}
and of the strong coupling (if they are colored); a large Yukawa or a large scalar quartic renormalize themselves to larger values at higher energy;
a large scalar cubic has the same problem, once vacuum stability bounds are taken into account~\cite{1602.01460}.

\smallskip

The Landau pole can be delayed or avoided if some extra comparably strong interaction is present.
For example, in the Standard Model 
the top Yukawa coupling $y_t\approx 1$ does not hit a nearby Landau pole
because the strong gauge coupling $g_3\sim y_t$ keeps the running $y_t$ under control,
providing a RGE flow with an infra-red fixed point.
Something similar can allow a larger $\F$ width, in models where
$N$ states with $q\sim 1$ lie in the fundamental of a new gauge group such as $\SU(N)$.~\cite{1602.01460}.

While it's premature to build models based on a $\sim1\sigma$ hint in favour of
a large width, models with a new strong interaction have been explored because of their own interest.

\section{Composite diphoton}
The above situation prompted many authors to consider strongly-coupled models, where the diphoton is a composite resonance.

\smallskip

Ref.s~\cite{1512.06670,1602.08100,1512.08221,1602.08819,1605.00013} explored the simplest possibility that $\F$ is a QCD bound state of heavy quarks $\bar{\Q}{\Q}$
with mass $M_\Q \approx \frac12 750\GeV$.
At this energy $\alpha_3 \approx 0.100$ is relatively small, such that the QCD binding energy is small:
only a small fraction of the produced $\bar{\Q}{\Q}$ pairs manifests as a 750 GeV resonance,
as well as inducing extra features in the $\gamma\gamma$ spectrum~\cite{1512.08221,1605.08741}.
The $\gamma\gamma$ excess can be reproduced if uncertain QCD factors are favourable,
if $\Q$ decays with a life-time longer than the life-time of the bound state,
and into a final state not subject to strong bounds.

\medskip

These difficulties can be alleviated  adding a new strong interaction
that confines just below $M_\Q$~\cite{big,1512.07733,1603.07719,1604.07828}.
In the limit where the  $\Q\bar\Q$ potential can be approximated as
$V(r)= - \alpha_{\rm eff}/r$ where $\alpha_{\rm eff}$ is some effective constant larger than $\alpha_3$,
the quarkonium-like resonances have mass
$M_n =2 M_\Q (1-\alpha_{\rm eff}^2/8n^2)$.
The lightest resonance with $n=1$ is identified with $\F$ and its decay widths are
\beq \label{eq:quarkonium}
\frac{\Ggg}{M_\F} = \frac{q^4 N }{4} \alpha_{\rm em}^2 \alpha_{\rm eff}^3=
10^{-6}  N q^4\bigg(\frac{\alpha_{\rm eff}}{0.4}\bigg)^3 ,\qquad
\Gamma_{gg}=\frac{2\alpha_3^2}{9q^4\alpha_{\rm em}^2}\Ggg
\eeq
having assumed that the particles $\Q$ are $N$ color triplets with charge $q$.
The slightly heavier resonances with $n>1$ have smaller widths, $\Ggg^{(n)}\propto 1/n^3$, giving a characteristic pattern.
However, since $3\otimes\bar 3 = 1\oplus 8$,
models based on a new strong interaction predict that each neutral resonance is accompanied by a quasi-degenerate
color octet resonance.
QCD repulsion reduces its binding energy, making its production cross sections less problematic.

%
%

\bigskip

Many authors explored the possibility that $\F$ is a bound state of a new  strong interaction.
There are three main classes of models:
\begin{enumerate}
\item {\bf Models where $H$ and $\F$ are composite}.~\cite{1602.06628,1605.04667}
This can be realised in simple fundamental models, where a new TechniColor (TC) gauge interaction
(for example with gauge group $\SU(N)$)
becomes strong around the weak scale,
and TC quarks are chiral under $\SU(2)_L$.
While the diphoton is totally natural, these models have big problems in reproducing higgs, electro-weak and flavor data.

\item {\bf Scenarios where $H$ and $\F$ are partially composite}~\cite{big,1512.07242,1512.05700,1601.05357,1605.09647} 
postulate chiral effective Lagrangians with the needed properties
that allow to bypass the TC problems, ignoring the issue of finding a fundamental dynamics that realises them.
The lightness of the Higgs is interpreted assuming that it is
the pseudo-Goldstone boson of an accidental global symmetry broken by unknown dynamics,
and often the pattern of symmetry breaking gives extra light singlets, that can be identified with $\F$.
Such models tend to give $\F\to t\bar t$ decays.

\item {\bf Models where $\F$ is composite}~\cite{1512.04850,1512.04924,big,1512.05334,1512.05759,1512.05779,1512.06827,1601.02490,1602.01092,1602.07297,1603.05774,1603.08802,1604.06180,1604.07776,1605.07183}
This can be realised in simple fundamental models where a new TC\footnote{Different authors use different names such as `dark', `hyper', `hidden', `big'  to distinguish the new gauge interaction 
from TechniColor. Since there is no unique name, we use the old-fashioned name Techni Color.}    gauge interaction
becomes strong around the weak scale, and
the TC particles are {\em not} chiral under the SM gauge group, that is thereby left unbroken by
the TC dynamics.
These models use an elementary Higgs doublet, like the SM, and are thereby equally compatible
with electro-weak, higgs and flavor data.\footnote{Before the announcement of the diphoton excess,
models of automatically stable composite Dark Matter based on such dynamics were explored,
even in papers that mentioned $pp\to\F\to\gamma\gamma$ signals.}
\end{enumerate}
Roughly speaking, the first class of models are dead, the second class are never born, so we focus on
the third class.
In order to obtain both $\F\to \gag$ and $gg\to\F$ the TC particles $\Q$ must be both colored and charged.
Then $\bar \Q\Q$ necessarily contains SM singlets, that can be identified with $\F$, but also color octets,
subject to strong LHC bound~\cite{1604.07835}.
If $M_\Q\circa{>}\Lambda_{\rm TC}$ such models realise the quarkonium-like scenario of eq.\eq{quarkonium},
which should be accompanied by a quasi-degenerate color octet.
More plausible realisations thereby identify $\F$ with a bound state that is much lighter than the others.
The main possibilities are:
\begin{itemize}

\item $\F$ as TC$\eta$.
If the TC dynamics breaks an accidental global symmetry, 
a set of $\bar\Q\Q$ pseudo-scalar bound states remains light, being the
TC-pions.  Some of them are neutral TC$\eta$.
Their couplings to SM vectors $V,V'$ are given in terms of the TC-pion decay constant $f_{\rm TC}$ and of
the gauge quantum numbers of the TC particles $\Q$
as
\beq
\frac{c_{VV'}}{\Lambda} = \frac{\kappa_{VV'} }{8\pi^2 f_{\rm TC}} ,\qquad\kappa_{VV'}= N \, {\rm Tr}\,(T_\F T^V T^{V'}).\eeq
where $T^V$ are the generators of the SM gauge group and
$T_\F$ is the chiral symmetry generator associated to $\F$
(for example $T_\F =\One/\sqrt{2N}$ corresponds to the TC$\eta'$ 
 state that gets a mass of order $\Lambda_{\rm TC}$ from TC anomalies).
So
\begin{equation}
\frac {\Ggg}{M_\F}=\frac {\alpha_{\rm em}^2}{64\pi^3}\frac { \kappa^2_{\gamma\gamma}  M_\X^2 }{f^2_{\rm TC}} = 
10^{-6}\bigg( \kappa_{\gamma\gamma} \frac{120\GeV}{f_{\rm TC}}\bigg)^2
\end{equation}
where $\kappa_{\gamma\gamma}=\kappa_{BB}+\kappa_{WW}$.
Measuring other $\F\to VV'$ decays would allow to infer the techni-particle content $\Q$.

\item{$\F$ as TC$\sigma$ or dilaton}.
Another state that can be especially light is the scalar
pseudo-Goldstone boson of scale invariance~\cite{big,1512.05330,1512.05618,1512.06106,1601.02570,1603.05668,1604.05328}.
Scale invariance is a good symmetry if two conditions are satisfied.
a) no TC particles have masses around $\Lambda_{\rm TC}$, unlike in QCD.
b) the TC-strong dynamics is in a `walking' regime, unlike the QCD dynamics where $\alpha_3$ `runs' to non-perturbative values.
These conditions imply an appropriate content of light TC particles $\Q$.

From a low-energy  perspective, the light TC-dilaton is the $\sigma$ field sometimes explicitly included in effective chiral Lagrangians~\cite{big,1603.05668}.
Its coupling to SM vectors is dictated by eq.\eq{LET} such that
\beq  \label{eq:techniGgg}
\frac{\Ggg}{M_\F} = 10^{-6} \bigg(\Delta b_{\rm em} \frac{120\GeV}{f_{\rm TC}}\bigg)^2 \eeq
where $ \Delta b_{\rm em}$ is the  contribution to the running of the electromagnetic coupling from techni-particles $\Q$.

\end{itemize}
Coming back to the issue of a large $\F$ total width, $\Gamma\sim 0.06M_\F$ 
can be realised adding extra $\F$ decays to SM particles or into other techni-pions,
which can include Dark Matter candidates.
However, a large $\Gamma$ needs a large $\Ggg\sim 10^{-3-4} M_\F$:
a look at all expressions for $\Ggg$ in composite $\F$ models
shows achieving such a large $\Ggg$  remains difficult.


\begin{table}
$$
\begin{array}{c|cccccccc}
&\multicolumn{6}{c}{\hbox{$\X$ couples to}}\\
& \multicolumn{6}{c}{\overbrace{\rule{8cm}{-1ex}}} \\
 \sqrt{s}=13\TeV & b\bar b& c\bar c & s\bar s & u\bar u & d\bar d & GG  \\ \hline 
\sigma_{\X  j}/\sigma_{\X} & 9.2 \% & 7.6 \% & 6.8 \% & 6.7 \% & 6.2 \% & 27. \%   \\
\sigma_{\X  b}/\sigma_{\X}  & 6.2 \% & 0 & 0 & 0 & 0 & 0.32 \%   \\
\sigma_{\X  jj}/\sigma_{\X}    & 1.4 \% & 1.0 \% & 0.95 \% & 1.2 \% & 1.0 \% & 4.7 \%  \\
\sigma_{\X  jb}/\sigma_{\X}  & 1.2 \% & 0.18 \% & 0.19 \% & 0.34 \% & 0.31 \% & 0.096 \%  \\
\sigma_{\X  bb}/\sigma_{\X}   & 0.31 \% & 0.17 \% & 0.18 \% & 0.34 \% & 0.31 \% & 0.024 \%   \\
\sigma_{\X  \gamma}/\sigma_{\X}    & 0.37 \% & 1.5 \% & 0.38 \% & 1.6 \% & 0.41 \% & \ll10^{-6}  \\
\sigma_{\X  Z}/\sigma_{\X}   & 1.1 \% & 1.1 \% & 1.3 \% & 2.0 \% & 1.9 \% & 3~10^{-6}  \\
\sigma_{\X  W^+}/\sigma_{\X}   & 5~10^{-5} & 1.7 \% & 2.4 \% & 2.6 \% & 4.1 \% & \ll10^{-6}   \\
\sigma_{\X  W^-}/\sigma_{\X}  & 3~10^{-5}  & 2.3 \% & 1.2 \% & 1.0 \% & 1.7 \% & \ll10^{-6}  \\
\sigma_{\X h}/\sigma_{\X}  & 1.0 \% & 1.1 \% & 1.2 \% & 1.9 \% & 1.8 \% &  1~10^{-6} 
\end{array}$$
\caption{\label{tab:sigmas}\em Predictions for the associated production of the resonance $\X$,
assuming the effective couplings of section~\ref{Leff}, whose validity is far from guaranteed.
We assumed the standard cuts 
$\eta_j<5$, $p_{Tj}>150\GeV$, $\Delta R_{jj}>0.4$ on jets,
and $\eta_{\gamma}<2.5$, $p_{T,\gamma}>10\GeV$ on photons.}
\end{table}

\section{What next?}\label{whatnext}
\begin{itemize}
\item[1)] Does $\F$ have spin 0, 2, or more?
\item[2)] Is $\F$ a $\SU(2)_L$ singlet or doublet or something else?
\item[3)] Is $\F$ produced through $gg$, $q\bar q$ or weak vector collisions?
\item[4)] Is $\F$ narrow or broad?   How large are its couplings and to which particles does $\F$  couple? 
\item[5)]  Is $\F$ CP-even or CP-odd or its couplings violate CP?
\item[6)] Is $\F$ elementary or composite?
\item[7)] Is $\F$ a cousin of the SM scalar?
\item[8)] Does $\F$ exist?
\end{itemize}
In the case of the SM scalar, these kind of questions had `a similar potential for surprise as a football game between Brazil and Tonga'.
In the case of $\F$ they are as open as a match between Brazil and Gemany.
Many  works explored how to ask these questions to nature and get answers~\cite{1512.06091,1512.06842,1512.07733,1601.03696,1602.05581,1602.07574,1604.02029,1604.04076,1604.06446,1604.06948,1605.05366,1605.00542,1605.05900,1605.07962,1606.02716,1606.03026,1606.03067,1607.00204,1607.01016}.
The main ideas are summarised below.

\begin{itemize}
\item[1)]
The {\bf spin} can be identified in the following ways:
\begin{itemize}
\item[$1a$)] spin 1 and half-integer spin are already excluded by the observation of $\F\to \gamma\gamma$.
\item[$1b$)] from angular distributions, as well known.
\item[$1c$)] a particle with spin 2 or higher can only be a bound state, that should come with other ones.
\end{itemize}
We will focus on spin 0.

\item[2)] The {\bf weak representation} of $\F$ can be identified in the following ways:
\begin{itemize}
\item[$2a$)] if $\F$ is not a neutral singlet, its extra charged components must be around 750 GeV. Find them.
\item[$2b$)] identifying the production mode: a singlet can couple at dimension 5 to all SM particles,
while a doublet is more likely to be produced from quarks, to which it may have renormalisable couplings.
\item[$2c$)] measuring the $p_T$ spectra in $\X$ associated production, in view of
the different dimensionalities of the effective couplings to quarks (dimension 5 for singlet $\X$ and 4 for doublet) and gauge bosons (dimension 5 for singlet and 6 for doublet).
\end{itemize}

\item[3)]
The {\bf initial state} that produces $\F$ can be identified in the following ways:
\begin{itemize}
\item[$3a$)] measuring how $\sigma(pp\to \F)$ depends on energy; data at $\sqrt{s}=8$ vs $13\TeV$ already disfavour production from light quarks or photons.
\item[$3b$)] any partonic $\wp\to\F$ production process implies a corresponding $\F\to\wp$ decay. Find it.
\item[$3c$)] the rapidity distribution and the transverse momentum spectrum of the diphoton system retain features of the initial parton state~\cite{1512.08478}.
\item[$3d$)] from the amount of extra jets from initial-state radiation in $pp\to\F$, see table~\ref{tab:sigmas}.
\item[$3e$)] production from $b$ quarks implies $\sigma(pp\to\X b) \approx6\%\,\sigma(pp\to\F)$ within the effective theory.
\item[$3f$)] $\X$ production in association with a gauge or Higgs boson is a useful discriminator, see table~\ref{tab:sigmas}. In particular, no vector bosons accompanying $\X$ are expected from gluon initial states, and no $W$ from $b$ initial states. Ratios of $\X W$, $\X Z$, and $\X h$ provide additional handles to identify the production process.  
\item[$3g$)] if $\F$ is a singlet produced from quarks, the $\X \bar q q H$ operator implies
a sizeable three-body decay width,
$\Gamma(\X\to q\bar q H) \sim 1\% \times \Gamma(\X\to q\bar q)$ where $H=\{h,Z,W^\pm\}$.
\end{itemize}

\item[4)]
The $\F$ {\bf couplings} can be measured in the following ways:
\begin{itemize}
\item[$4a$)] If $\F$ is broad enough that its total width can be measured, such that the couplings
can be reconstructed from the branching ratios.
\item[$4b$)]  If $\F$ is broad, it might even be possible to measure interference with SM cross sections~\cite{1601.00006,1604.06446,1605.00542,1605.04944,1606.03026}.
\item[$4c$)] $\SU(2)_L$-invariance relates different decay widths, allowing to disentangle the effective operators.
\item[$4d$)] $\F$ couplings to DM can be accessed from missing energy signals in the usual ways.
\item[$4e$)] Associated processes such as $pp\to \F j$ or $pp\to \F V$ probe the energy dependence of the couplings.
\item[$4f$)] Observation of $pp\to \F\F$,
would imply relatively large couplings:
either $\F^3$ cubics, or of $\F$ to the particles $\Q$ that mediate $\F\to\gamma\gamma$
(this effect can be computed by expanding eq.\eq{LET} to order $\F^2$):
$$\raisebox{-1.5cm}{ \includegraphics[height=3cm]{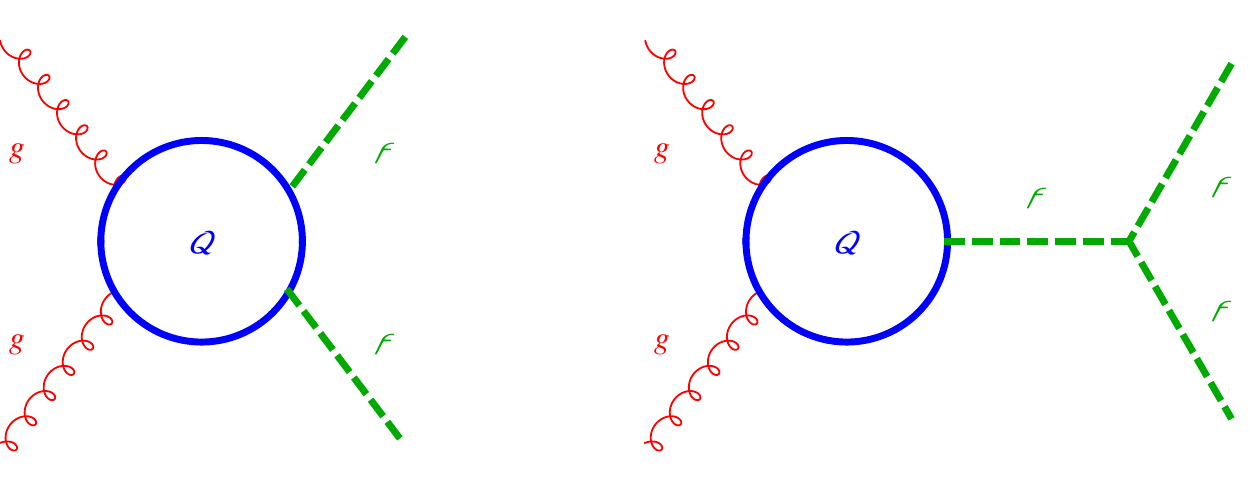}}\label{eq:FeynSS}$$
\end{itemize}

\item[5)]
The {\bf CP parity} of $\F$ can be identified in the following ways:
\begin{itemize}
\item[$5a$)] From angular distributions of $\X \to\gamma^* \gamma^*\to \ell^+\ell^- \ell^{\prime+}\ell^{\prime -}$; however its rate is 3 orders of magnitude below
the $\gamma\gamma$ rate.
\item[$5b$)] From $\F\to\gamma\gamma$ using $\gamma\to e^+e^-$ conversions in the detector matter; however the angle between the $e^\pm$ pairs is very small.
\item[$5c$)] From the angular distribution of $pp\to jj\F$ events.
\item[$5d$)] If $\F\to ZZ$ exists, the CP parity of $\F$ can be measured from angular distributions in leptonic $Z$ decays and in $pp\to Z\F$.
\item[$5e$)] If $\F\to Z\gamma$ exists, something can be done combining $5a$) with $5d$).
\item[$5f$)] If $\F\to hh$ exists, it implies that $\F$ is a scalar.
\item[$5g$)] If $\F\to hZ$ exists, it implies that $\F$ is a pseudo-scalar, and that the effective theory approximation fails, given that low-dimension operators
do not induce such decay.
\end{itemize}
\end{itemize}

\section{Connection with Dark Matter, axions, vacuum stability, baryogenesis...}\label{DM}
Possible connections of $\F$ with other open issues have been investigated.

Various authors explored the possibility that the diphoton is the mediator that couples  Dark Matter to SM particles~\cite{1512.04913,1512.04917,1512.06376,1512.06787,1512.07243,1512.08117,1512.08992,1601.01381,1601.05038,1603.04479,1603.05601,1604.05774,1605.04804,1605.05327,1605.09018,1606.00557}.
The freeze-out DM relic abundance can reproduce the observed
cosmological DM abundance for natural values of the parameters,
as it is customary for weak-scale particles.
In particular, if the diphoton decays into Dark Matter,
one needs a DM mass around 100-300 GeV, depending on the diphoton width.
DM direct detection is somewhat below present bounds if the diphoton is a scalar,
or suppressed by non-relativistic factors if the diphoton is a pseudo-scalar.

\medskip

Adding to the SM field an axion allows to understand the smallness of the CP-violating $\theta$ angle of QCD.
However the axion must be ultra-light and coupled ultra-weakly: it cannot be identified with $\F$.
Nevertheless, some authors tried to identify the $\F$ resonance with an axion-like state, by building models where it can be heavy and significantly coupled~\cite{1512.04931,1512.08467,1512.08777,1602.00949,1602.07909,1604.01127,1606.03097}.
For example~\cite{1604.01127} tries to realise a Nelson-Barr-like model at the weak scale.

\medskip

The addition of $\F$ can eliminate the  instability of the SM vacuum~\cite{1512.06782,1512.07889,1512.08184,1605.08681} in two different ways:
a) by providing a tree-level threshold corrections that increases the Higgs quartic $\lambda_H$;
b) the extra charged particles that mediate $\F\to\gamma\gamma$ modify the RGEs,
such that $g_{2,Y}$ and consequently $\lambda_H$ become larger at large energy.

\medskip

The electro-weak  phase transition, extended including $\F$, 
could become of first order leading to gravitational waves and baryogengesis~\cite{1601.04291,1601.04678,1603.04488,1604.02382,1605.00037,1605.08681,1605.08736}.
Other works discussed connections with neutrino masses~\cite{1601.00386,1601.02714,1601.05038,1603.07672,1606.05163},
 flavor~\cite{1512.06828,1512.08508,1512.09092,1604.03940,1606.07082},
 inflation~\cite{1512.06782,1512.08984,1512.09136,1606.06677},
 extra dimensions~\cite{1512.05771,1512.06335,1512.08440,1512.06674,1512.07672,1601.07167,1602.02793,1603.04495,1603.06980,1603.07303,1603.09550,1607.01464},
and string/$\F$-theory~\cite{1512.06773,
1512.07622,
1512.08502,
1512.08777,
1601.00640,
1601.00285,
1601.03604,
1602.09099,1603.06962,1603.08294,1606.01052,1606.02956,1607.04534}.

\section{Who ordered that?}
After that experiments will answer the questions of section~\ref{whatnext}, 
clarifying what $\F$ really is,
it will be possible to understand which role $\F$ plays in particle physics.

In the meantime, various authors started to explore the possibility that $\F$ has something to do with the origin of the electro-weak scale,
trying to identify $\F$ with one or another supersymmetric particle:
sneutrino~\cite{big,1512.06560,1512.07645}, extra scalar or pseudo-scalar Higgs~\cite{big,1512.08434,1512.09127,1605.01040,1606.04131},  extra NMSSM singlet~\cite{big,1603.00718,1512.08323,1512.09127,1601.00866,1601.07242,1602.03344,1604.03598}, sgoldstino~\cite{1512.05330,1512.05333,1512.05723,1512.07895,1602.00977,1603.05251,1603.05682}
(its rate in $\gamma\gamma$ implied by gaugino masses seems too small), stopponia~\cite{1605.00013,1606.08811},
sbino~\cite{1512.06107,1605.05313} or else~\cite{1512.06696,1512.07904,1601.00633,1602.07866}.
In this context, extra full SU(5) multiples around the weak scale can enhance the $\gamma\gamma$ rate~\cite{1512.07468,1512.07904,1601.00866,1604.03598,1605.03585,1606.08785}.
Connections with other solutions to the hierarchy problem have been also explored, such as
composite or partially composite Higgs or extra dimensions.

A related theoretical issue is the naturalness of the extra charged particles introduced to mediate $\F\to\gamma\gamma$.
Even if they are fermions, they have no chirality reason to be around the weak scale, unlike the SM fermions.\footnote{Unless an enlarged gauge symmetry broken around the weak scale, 
such as $\SU(3)_L\otimes{\rm U}(1)\otimes\SU(3)_c $~\cite{1512.06878,1512.07165,1512.08441,1606.03415} or $\SU(3)_L\otimes\SU(3)_R\otimes\SU(3)_c$~\cite{1512.07225}
or $\SU(6)\to G_{\rm SM}\otimes\U(1)$~\cite{1604.07838},
provides an extended set of charged chiral fermions. Furthermore, $\SU(3+N)\to\SU(3)_c$ provides extra charged scalars~\cite{1606.05865}.
}
Supersymmetry and other solutions to the hierarchy problem imply extra charged particles at the weak scale.
Indeed, such extensions of the SM tame quadratically divergent corrections to the Higgs mass at the price of
introducing a lot of new physics at the weak scale.

However, such new physics has not been observed, and
bounds relegate solutions to the hierarchy problem to fine-tuned corners of their parameter space.
Some authors, following the point of view that quadratic divergences give no physical effects,
explored models where no physical correction is unnaturally large 
and tried to build extensions of the SM where a hierarchically small weak scale is induced by some new dynamics.
Ref.s.~\cite{1512.06708,1512.07225,1512.09136,1605.08681} explored the possibility that $\F$ is (a manifestation of) such dynamics:
the smoking gun of this scenario would be observing that $\F$ couples to all particles proportionally to their masses.
In this context, broken scale invariance can justify extra charged particles, which cannot be much above the weak scale
in order to avoid unnaturally large physical corrections to the Higgs mass.

%

\section{Conclusions}\noindent
The $\F\to\gamma\gamma$ excess can be a statistical fluctuation, or a the first sign of new physics.
In the latter case various reasonable models can reproduce it, at least if the $\F$ width is narrow.
A generic prediction is that we expect to see also
$\F\to\gamma Z$ and/or $\F\to ZZ$ and extra charged particles.
All the rest is model-dependent.
Today the diphoton excess could be everything, including nothing.

\paragraph{Note Added}
The above discussion is a disproportionate amplification of a quantum  fluctuation:
no 750 GeV $\gamma\gamma$ excess is present in the first $(12.2+12.9)\,{\rm fb}^{-1}$ of new 2016 LHC data~\cite{newdata}, 
which confirm the Standardissimo Model and the
bad reputation of the digamma symbol $\F$.

%
%
%



\small

\section*{Acknowledgments}
I thank the organisers for creating a  unique environment for  interchange of ideas,  all participants for a multitude of new insights.
This work was supported by the ERC-AdG-2014 grant 669668 -- NEO-NAT.

\end{document}